\documentclass[runningheads,citeauthoryear]{apinv_kurtedit}
\usepackage{epsfig,cite,graphics}
\usepackage{marvosym} 
\usepackage[utf8]{inputenc}
\usepackage{longtable}
\newcommand\TT{\rule{0pt}{1.9ex}}

\begin{document}

\title{Searching for twins of the V1309 Sco progenitor system:\\ a selection of long-period contact binaries}
\titlerunning{Searching for twins of the V1309 Sco progenitor system}
\author{Alexander Kurtenkov\inst{1,2}}
\authorrunning{A. Kurtenkov}
\tocauthor{A. Kurtenkov} 
\institute{Department of Astronomy, University of Sofia, 5 James Bourchier Blvd., 1164 Sofia, Bulgaria
	\and Institute of Astronomy and National Astronomical Observatory, Bulgarian Academy of Sciences, 72 Tsarigradsko Shose Blvd., 1784 Sofia, Bulgaria \\
	\email{al.kurtenkov@gmail.com}    }
\papertype{Submitted on xx.xx.xxxx; Accepted on xx.xx.xxxx}	
\maketitle

\begin{abstract}
The only well-studied red nova progenitor (V1309 Sco) was a contact binary with a 1.4-day period. The prospects for searching for similar systems, as well as stellar merger candidates in general, are explored in this work. The photospheric temperatures of 128 variables with periods $P=1.1-1.8$\,d classified as W UMa-type binaries are calculated using their colors listed in the SDSS catalog. A selection of 15 contact binaries with similar temperatures and periods as the V1309 Sco progenitor is thus compiled. The Kepler Eclipsing Binary Catalog is used to analyse systems with eclipse timing variations (ETV) possibly caused by changes of the orbital period. Out of the 31 systems with parabolic ETV curves listed by Conroy et al. (2014, AJ, 147, 45) two could be contact binaries with a decreasing period and, therefore, potential stellar merger candidates. Out of the 569 contact binaries in the OGLE field analysed by Kubiak et al. (2006, AcA, 56, 253) 14 systems have periods longer than 0.8\,d and a statistically significant period decrease.
\end{abstract}
\keywords{Stars: binaries: eclipsing -- Stars: individual: V1309 Sco -- Stars: evolution}

\section*{1. Introduction}
The outburst of V1309 Sco in 2008 has played a key role in our understanding of the rare red nova eruptions. Very few such events were previously known. The well-studied V838 Mon (Bond et al. 2003) had shown distinct similarities in its spectral evolution to transients V4332 Sgr and M31 RV. Tylenda \& Soker (2006) suggested that all three events were caused by mergers in binary stars. However, it was not until the eruption of V1309 Sco that direct observations of a red nova progenitor could support this theory.

Mason et al. (2010) presented thorough observations of the V1309 Sco outburst and pointed out its similarities to red novae. They derived a peak absolute magnitude of $M_{V}=-8.3$\,mag and a maximum emission line FWHM of $\sim150$\,km/s. Tylenda et al. (2011) used archive data from the OGLE project to show that the progenitor is a W UMa (contact) binary with a period of $1.44$\,d and a $\sim0.2$\,mag amplitude. Moreover, the period had been decreasing exponentially for at least six years before the outburst, reaching $1.42$\,d in 2007. The OGLE lightcurve also showed slow brightness variations as the I magnitude remained in the range $16-17$\,mag before the outburst. Tylenda et al. (2011) obtained an effective temperature of the progenitor of 4500\,K under certain approximations for the interstellar reddening. An exact value was difficult to derive due to the proximity of V1309 Sco to the galactic plane. 

Five galactic red novae are currently known -- V4332 Sgr, V838 Mon, V1309 Sco, the highly reddened OGLE-2002-BLG-360 and allegedly, the 1670 eruption of CK Vul (Kaminski et al. 2015). With the addition of several extragalactic transients, such as M31 RV (Rich et al. 1989), M85 OT2006-1 (Kulkarni et al. 2007), and M31LRN 2015 (Kurtenkov et al. 2015), they constitute a rare class of mostly very luminous events. As of 2016 V1309 Sco remains the only red nova with a well-studied progenitor, and the only direct proof that stellar mergers cause red novae.

\section*{2. Motivation}

The archival observations of V1309 Sco have presented the unique opportunity of studying the transition of a stellar merger process from a quiescent phase (eclipsing binary with a decreasing orbital period) to a dynamic phase (red nova explosion, ultimately leaving the merger product engulfed in dust). A prediction of such an event will allow for a much more detailed exploration of the progenitor system, e.g. mass and temperature determination via multicolor photometry, or observing the spectral evolution of the common envelope. Such results can be used to improve the relations between the parameters of the progenitor system and the parameters of the transient. An attempt at studying these relations was made by Kochanek et al. (2014), although the statistical sample is still quite small. A red nova prediction was made by Molnar et al. (2015) for the Kepler eclipsing binary KIC 9832227.

As discussed by Tylenda et al. (2011) the merging can be triggered by the evolution of the primary component. It grows in size as it crosses the Hertzsprung gap and the system starts losing mass and angular momentum through the $L_{2}$ Lagrangian point. As the V1309 Sco progenitor probably included an evolved K-star, this scenario is compatible with the observations. Furthermore, the observed orbital period of $1.44$\,d is unusually long for W UMa-type binaries, so it requires a large semi-major axis, and respectively, large radii. Similarly, many long-period contact binaries can be seen as potential merger candidates and their exploration could give us an insight into the physics of red nova progenitors.

The last decades have shown a radical increase in the number of known eclipsing binary systems, owing to modern time-based sky surveys, such as NSVS, ASAS, and most recently, the Catalina Sky Survey. The latter has produced a catalog including $\sim31000$ contact and ellipsoidal binaries (Drake et al. 2014). As of June 2016 the VSX (Variable Star Index) database lists a total of 39634 W UMa-type binaries. The present work aims to show that the number of known contact binaries is sufficient to draw an object selection of possible V1309 Sco-like systems. Contact binaries that show similar characteristics as V1309 Sco could be at the same evolutionary stage at which the mechanism that led to the 2008 eruption is triggered. The selected systems in Section 3 could prove to be promising targets of follow-up observations, aiming to derive their absolute parameters and to look for period changes. 

Another important addition to the exploration of eclipsing binaries has been the Kepler Eclipsing Binary Catalog (Slawson et al. 2011). It could be a powerful tool in the search of stellar merger candidates as it provides high-quality lightcurves with exact minima timing. Section 4 is dedicated to selecting stellar merger candidates from Kepler and OGLE binaries with considerable eclipse timing variations.  

\section*{3. SDSS/VSX color-based selection}

\begin{figure}
\centering
\includegraphics[width=12.9cm]{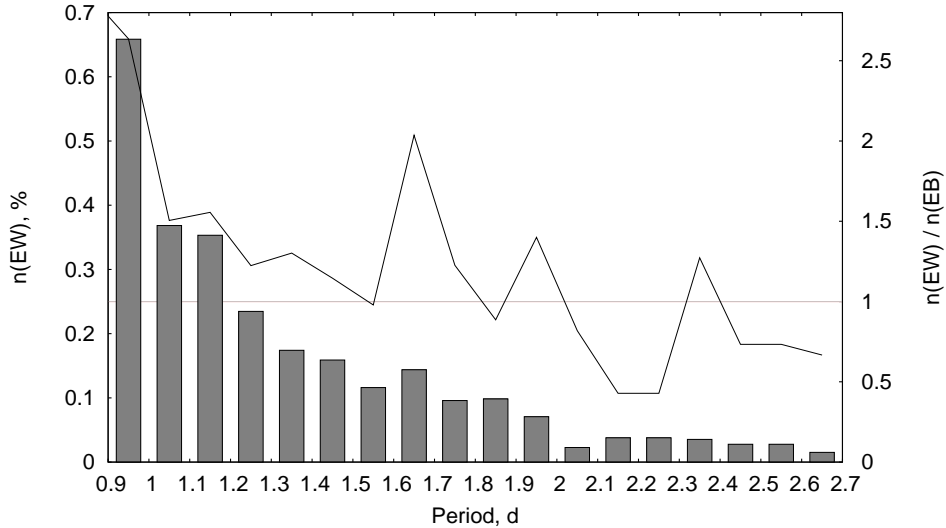}
\caption{The bars represent the period distribution of systems classified as W UMa binaries for $P>0.9$\,d, expressed in percentages of all 39634 such systems in the VSX. The dark line represents the ratio of cataloged W UMa to $\beta$ Lyr binaries for each period interval. The ratio decreases after $P\sim2.0$\,d, which implies that the majority of cataloged W UMa binaries with periods in the $1.0-2.0$\,d range are not misclassified $\beta$ Lyrae. The left y-scale applies to bars, while the right y-scale applies to lines. All data is extracted from the VSX database.}
\label{P_distribution}
\end{figure}

Binaries of the W UMa type have typical orbital periods in the range $0.2-0.8$\,d. However, the VSX database currently lists 1125 W UMa binaries with $P>1$\,d. Some of them are misclassified $\beta$ Lyrae as those variables have similar lightcurve shapes but generally larger periods. The distribution of cataloged long-period W UMa binaries is shown in Fig.\,\ref{P_distribution}. In the $1.0-2.0$\,d period range there are still relatively many systems cataloged as W UMa binaries. Also plotted is how the ratio of cataloged W UMa to $\beta$ Lyr binaries changes with the orbital period. In case all systems in the $1.0-2.0$\,d period range cataloged as contact binaries were misclassified $\beta$ Lyrae, this ratio would not decrease after $P\sim2.0$\,d. Many systems in this period range could be contact binaries with one or two evolved components and, therefore, objects of the current study.

The somewhat narrower period range between 1.1\,d and 1.8\,d is approximately centered on the V1309 Sco period. As of June 2016 the VSX lists 506 systems of the W UMa type in it. Of those, 207 were identified in the ninth release of the SDSS catalog (Ahn et al. 2012) within a search radius of 2\arcsec. The SDSS catalog provides multicolor photometry in the Sloan $ugriz$ passbands. The imaging data of a given object is obtained with a $72$\,sec delay between two consecutive filters, so there is a $\sim5$\,min interval between the $r$-band and $g$-band frames. Therefore it can be safely assumed that the imaging of $P>1d$ binaries in all passbands is simultaneous. The colors of contact binaries slightly vary with phase due to the temperature difference of the components and the gravity darkening effect. Both effects are of the order of $\sim10^{2}$\,K, which allows for a satisfactory estimate of the temperature of the common envelope using multicolor photometry at a given phase. The color-temperature relations derived by Boyajian et al. (2013) were used for this purpose. These relations are of the form

\begin{eqnarray}
    T_{eff} &=& a_{0}+a_{1}X+a_{2}X^{2}+a_{3}X^{3},                           
\end{eqnarray}
where X represents the color index and $T_{eff}$ is the effective photospheric temperature. The coefficients derived by Boyajian et al. (2013) for the Sloan photometric system are given in Tab.\,\ref{color-T}.

\begin{table}
\begin{center}
\caption[Polynomial coefficients of the color-temperature relation extracted from Boyajian et al. (2013).]{Polynomial coefficients of the color-temperature relation extracted from Boyajian et al. (2013).}\label{color-T}
\small
\begin{tabular}{c@{ }c@{ }c@{ }c@{ }c@{ }c@{ }}
\hline\hline
\TT Color index \,& $a_{0}$  & $a_{1}$ & $a_{2}$ & $a_{3}$ & Range [mag] \\
\hline
\TT g-r	\,\,& 7526	\,\,& $-5570  $ \,\,& $3750$ \,\,& $ -1332.9$	\,\,& $-0.23-1.40 $ \\
\TT g-i	\,\,& 7279	\,\,& $-3356  $ \,\,& $1112$ \,\,& $ -153.9$	\,\,& $-0.43-2.78 $ \\
\TT g-z	\,\,& 7089	\,\,& $-2760  $ \,\,& $804$  \,\,& $ -95.2$	\,\,& $-0.58-3.44 $ \\
\hline        
\end{tabular} 
\end{center}  
\end{table}  

\begin{table}
\begin{center}
\caption[A selection of 23 contact binaries with $P=1.1-1.8$\,d and $T_{eff}=4200-4800$\,K. The temperatures of the systems correspond to spectral types K0III-K3III and are similar to the V1309 Sco progenitor.]{A selection of 23 contact binaries with $P=1.1-1.8$\,d and $T_{eff}=4200-4800$\,K. The temperatures of the systems correspond to spectral types K0III-K3III and are similar to the V1309 Sco progenitor. The last column contains the values of the total galactic reddening along the line-of-sight (the maximum possible reddening of the object). The visual (wide band V) magnitudes and amplitudes are extracted from the VSX database.}\label{results}
\small
\begin{tabular}{c@{ }c@{ }c@{ }c@{ }c@{ }c@{ }c@{ }c@{ }c@{ }}
\hline\hline
\TT $\alpha$(J2000.0) \,\,& $\delta$(J2000.0)  \,\,& P [d] \,\,& Mag. \,\,& Amp. \,\,& SDSS epoch \,\,& $T_{eff}$[K] & $\sigma_{T}$[K] & $E(B-V)_{0}$ \\
\hline \TT   
   10:00:01.70 \,&  24:23:05.1  \,&   1.2955 \,&  15.85 \,&  0.15  \,&   2004.9566  \,&   4213   \,&   48  \,&    0.03     \\
   13:06:00.14 \,&  18:00:46.7  \,&   1.3074 \,&  16.68 \,&  0.14  \,&   2005.4271  \,&   4315   \,&   70  \,&    0.02     \\
   05:44:53.66 \,&  03:18:35.0  \,&   1.1997 \,&  16.17 \,&  0.18  \,&   2007.8835  \,&   4365   \,&  100  \,&    0.75     \\
   16:56:09.74 \,&  42:42:19.7  \,&   1.2699 \,&  15.52 \,&  0.18  \,&   2004.4527  \,&   4383   \,&  127  \,&    0.02     \\
   03:29:54.17 \,&  38:01:44.9  \,&   1.3618 \,&  14.22 \,&  0.25  \,&   2003.0859  \,&   4398   \,&   66  \,&    0.27     \\
   08:31:48.57 \,&  29:44:44.0  \,&   1.4136 \,&  14.93 \,&  0.12  \,&   2002.9988  \,&   4445   \,&   89  \,&    0.03     \\
   05:44:07.39 \,&  28:31:40.8  \,&   1.2753 \,&  15.77 \,&  0.78  \,&   2003.9705  \,&   4456   \,&  130  \,&    1.26     \\
   17:14:28.01 \,&  10:13:10.7  \,&   1.7623 \,&  15.92 \,&  0.14  \,&   2005.4380  \,&   4479   \,&   61  \,&    0.10     \\
   02:09:03.05 \,& -04:26:14.7  \,&   1.5781 \,&  18.17 \,&  0.31  \,&   2005.7421  \,&   4496   \,&  116  \,&    0.02     \\
   08:17:46.29 \,&  13:05:05.7  \,&   1.5947 \,&  14.90 \,&  0.08  \,&   2004.9704  \,&   4519   \,&   64  \,&    0.02     \\
   07:34:32.94 \,&  36:48:28.5  \,&   1.1646 \,&  17.38 \,&  0.31  \,&   2000.2604  \,&   4519   \,&   96  \,&    0.05     \\
   05:07:27.74 \,&  16:13:10.5  \,&   1.1861 \,&  15.21 \,&  0.23  \,&   2006.8322  \,&   4528   \,&   60  \,&    0.31     \\
   22:29:09.15 \,&  17:37:23.4  \,&   1.3664 \,&  14.60 \,&  0.08  \,&   2009.7912  \,&   4568   \,&   46  \,&    0.06     \\
   14:21:14.77 \,&  35:28:38.9  \,&   1.5851 \,&  16.14 \,&  0.09  \,&   2003.3164  \,&   4592   \,&   26  \,&    0.02     \\
   04:25:51.08 \,&  30:02:48.9  \,&   1.1489 \,&  15.72 \,&  0.38  \,&   2002.9984  \,&   4613   \,&  100  \,&    0.39     \\
   04:39:40.76 \,&  19:18:53.3  \,&   1.7422 \,&  13.92 \,&  0.30  \,&   2006.0841  \,&   4658   \,&   99  \,&    0.34     \\
   23:15:47.32 \,&  12:29:37.5  \,&   1.7325 \,&  15.28 \,&  0.19  \,&   2008.8385  \,&   4679   \,&   75  \,&    0.06     \\
   04:45:50.71 \,&  11:34:57.2  \,&   1.3164 \,&  15.29 \,&  0.35  \,&   2006.8321  \,&   4723   \,&  110  \,&    0.51     \\
   15:02:20.15 \,&  26:21:12.1  \,&   1.7954 \,&  15.94 \,&  0.18  \,&   2004.3649  \,&   4729   \,&   83  \,&    0.04     \\
   07:26:18.08 \,&  31:55:03.0  \,&   1.1579 \,&  15.71 \,&  0.22  \,&   2006.8870  \,&   4744   \,&   67  \,&    0.06     \\
   09:46:14.73 \,&  16:47:06.0  \,&   1.6007 \,&  15.31 \,&  0.25  \,&   2005.1943  \,&   4750   \,&  100  \,&    0.03     \\
   00:44:12.88 \,&  20:11:07.8  \,&   1.7533 \,&  13.75 \,&  0.06  \,&   2009.0463  \,&   4750   \,&   77  \,&    0.04     \\
   10:02:30.35 \,&  50:18:08.8  \,&   1.2404 \,&  15.51 \,&  0.23  \,&   2001.8904  \,&   4760   \,&   74  \,&    0.01     \\
\hline        
\end{tabular} 
\end{center}  
\end{table}  

Three color temperatures ($T_{g-r}$, $T_{g-i}$ and $T_{g-z}$) were calculated for each of the 207 objects. The photospheric temperature $T_{eff}$ and its statistical standard error $\sigma_{T}$ were then calculated as an average of the three values and the RMS of the residuals, respectively. Photospheric temperatures for 143 systems were thus obtained. The color indices of 9 objects are outside the calibration range. The remaining 55 objects (26\,\% of all) have $\sigma_{T}>150$\,K and were excluded from the sample. The results are in the range $3230-7560$\,K (median temperature 5150\,K), with a median error $\sigma_{T}=80$\,K.

The 23 systems with $T_{eff}=4200-4800$\,K are listed in Tab.\,\ref{results}. Giants of the K0III-K3III types have effective temperatures in this range. It should be noted that the interstellar reddening has been neglected for this calculation, so it is plausible only for objects that are far from the galactic plane, where the reddening is weak. For 15 of the 23 systems the total galactic reddening along the line-of-sight is less than 0.07\,mag. If these are correctly classified as W UMa binaries, they may have similar parameters to the V1309 Sco progenitor. Follow-up observations of these systems are therefore encouraged.

Of the 143 systems 128 have galactic latitudes $|b|>20$\degr, which implies a low reddening. The calculated photospheric temperatures for all 128 systems are presented in the online Tab.\,\ref{all}.

\section*{4. Kepler and OGLE candidates}

The objects selected in Section 3 might be of a similar evolutionary stage as the progenitor of V1309 Sco, but it is highly improbable that they will erupt in the upcoming decades. The best way to predict such an event is based on minima timing as the decrease of the orbital period before merging causes eclipse timing variations (ETV). The most reliable source of such information is by far the Kepler mission.

\begin{table}
\begin{center}
\caption[.]{Kepler systems classified after visual inspection. The following designations are used: EA = Algol, EB = $\beta$ Lyr, EW = W UMa, ELV = ellipsoidal variable; Pos = positive, Neg = negative.}\label{table_etv}
\small
\begin{tabular}{c@{ }c@{ }c@{ }c@{ }c@{ }c@{ }c@{ }c@{ }}
\hline\hline
KIC	\,&	P [d] \,& type \,& $d^{2}ETV/dt^{2}$ \,&	KIC	\,&	P [d] \,& type \,& $d^{2}ETV/dt^{2}$ \\
\hline
2305372   \,&	1.404	\,& EA       \,&  Pos	\,& 6677225   \,&   0.525   \,& ELV/EW   \,&  Pos	   \\
3104113   \,&	0.846	\,& EW       \,&  Pos	\,& 7696778   \,&   0.331   \,& EW	 \,&  Pos	   \\
3765708	  \,&	0.431	\,& ELV/EW   \,&  Pos	\,& 7938468   \,&   7.226   \,& EA	 \,&  Pos	   \\
4074532   \,&	0.353	\,& EW       \,&  Pos	\,& 7938870   \,&   0.580   \,& EA/EB	 \,&  Neg	   \\
4851217   \,&	2.470	\,& EA       \,&  Pos	\,& 8758161   \,&   0.998   \,& EA	 \,&  periodic?    \\
4853067	  \,&	1.340	\,& ELV/EB   \,&  Neg	\,& 9087918   \,&   0.445   \,& EW	 \,&  Pos	   \\
5020034   \,&	2.119	\,& EA       \,&  Pos	\,& 9402652   \,&   1.073   \,& EA	 \,&  periodic?    \\
5471619   \,&	0.962	\,& EA/EB    \,&  Neg	\,& 9840412   \,&   0.878   \,& EW	 \,&  Neg	   \\
5770860	  \,&	0.737	\,& ELV/EW   \,&  Pos	\,& 9934052   \,&   0.352   \,& ELV/EW   \,&  Pos	   \\
5792093   \,&	0.600	\,& EB       \,&  Neg	\,& 10030943  \,&   0.235   \,& ELV/EW   \,&  Pos	   \\
6044064   \,&	5.063	\,& EA       \,&  Pos	\,& 10292413  \,&   0.559   \,& ELV/EW   \,&  Neg	   \\
6044543	  \,&	0.532	\,& EB       \,&  Neg	\,& 10736223  \,&   1.105   \,& EA	 \,&  Pos	   \\
6066379   \,&	1.303	\,& EA       \,&  Neg	\,& 11097678  \,&   0.999   \,& EW	 \,&  Pos	   \\
6213131   \,&	0.561	\,& EB       \,&  Pos	\,& 11144556  \,&   0.642   \,& EW	 \,&  Pos	   \\
6314173	  \,&	1.433	\,& EA       \,&  Neg	\,& 11924311  \,&   0.445   \,& EB/EW	 \,&  Pos	   \\
6464285   \,&	0.843	\,& EA       \,&  Pos	\,& ...     \,&   ...	\,& ...    \,&  ...		   \\
\hline
\end{tabular} 
\end{center}  
\end{table} 

The presence of ETV does not necessarily imply a loss of angular momentum from the system. It is far more likely to be caused by mass transfer between the components or by signal delay due to perturbations by a third body (light travel time effect, LTTE). In the latter case the ETV curve exhibits strict periodicity. In the former case, which is quite common for $\beta$ Lyrae, the second derivative of the ETV curve can be positive as well as negative. Only contact binaries with parabolic ETV curves with a negative second derivative could undergo a stellar merger in the near future.

\begin{figure}[!ht]
\centering
\includegraphics[width=12.9cm]{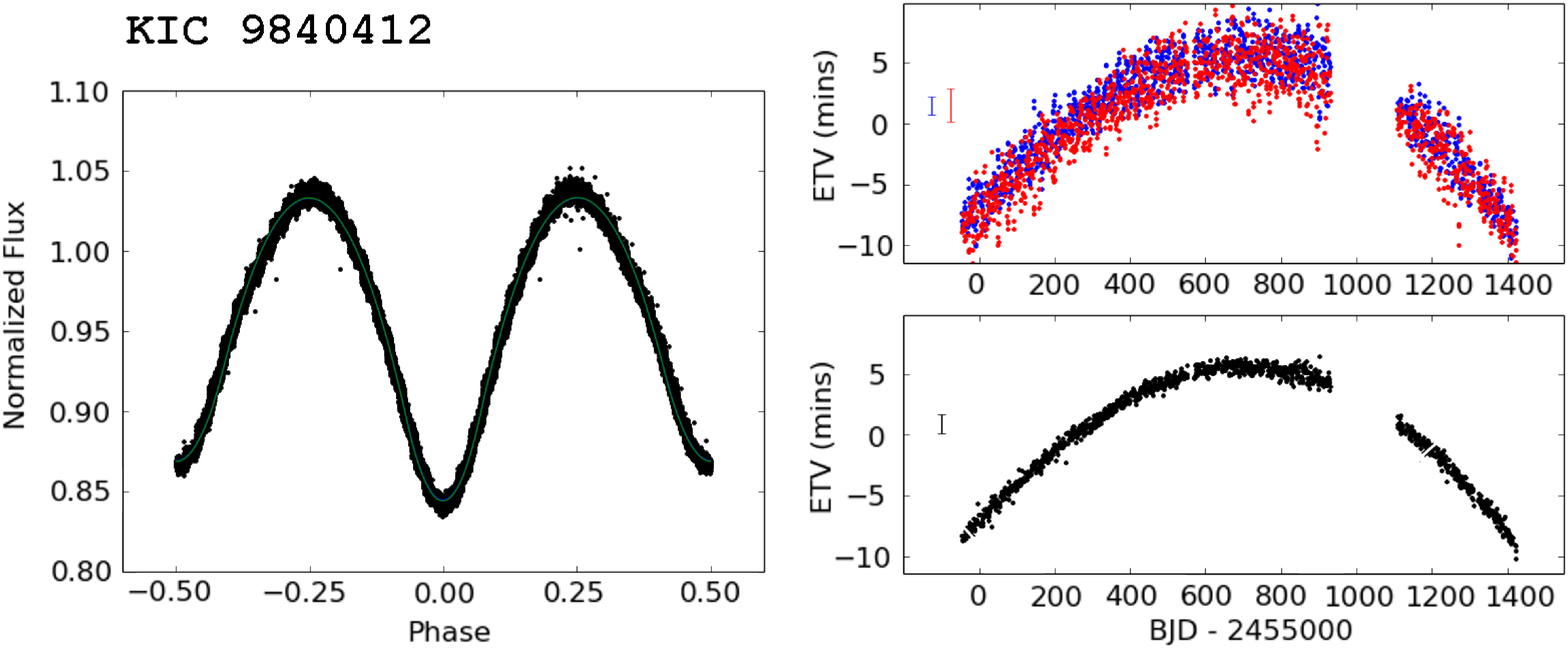}
\includegraphics[width=12.9cm]{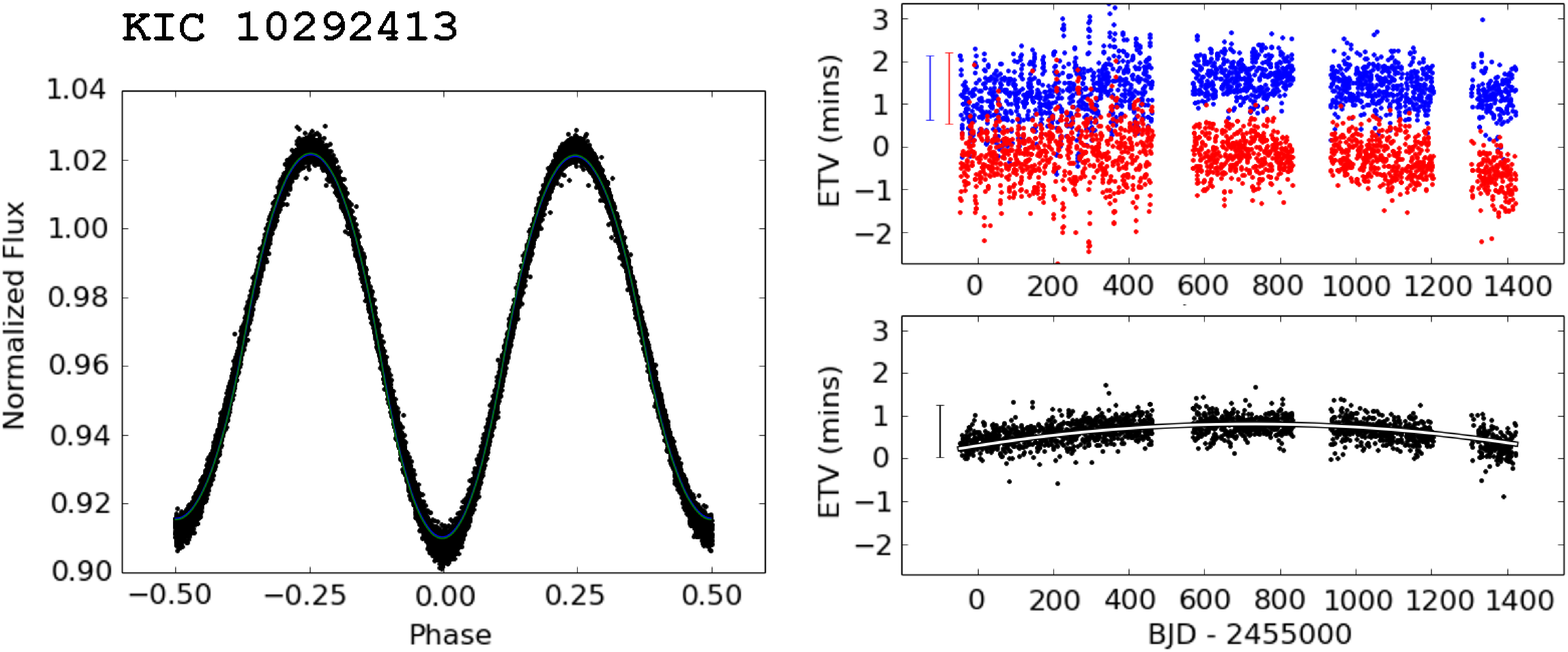}
\caption{Lightcurves (\textit{left}) and ETV curves (\textit{right}) of the two selected W UMa binaries in the Kepler field. Both ETV curves (\textit{top}: of both the primary and secondary minima, \textit{bottom}: averaged) have a negative second derivative. The images are retrieved from the Kepler Eclipsing Binaries database at \url{http://keplerebs.villanova.edu/}.}
\label{fig_etv}
\end{figure}

The ETV signals of 1279 Kepler binaries were analysed by Conroy et al. (2014). The authors found 31 systems with nearly parabolic ETV curves over the whole $\sim1400$\,d period of observation. The type of variability was determined for all 31 systems after visual inspection of the lightcurves and ETV curves (Tab.\,\ref{table_etv}). Just two contact binaries with a negative second derivative of the ETV curve were found: 
\begin{itemize}  
\item KIC 9840412 ($BJD_{0}=2454954.556807$, $P=0.87845(58\pm9)$\,d) 
\item KIC 10292413 ($BJD_{0}=2454954.202209$, $P=0.55915(88\pm4)$\,d). \\
\end{itemize}

The latter could as well be an ellipsoidal variable. Note that ellipsoidal variables are hard to distinguish from contact binaries with nearly sinusoidal lightcurves. The lightcurves and ETV curves of the two systems are presented in Fig.\,\ref{fig_etv}. The period derivatives were calculated from $\frac{dP}{dt}=P\frac{d^{2}ETV}{dt^{2}}$. The parabolic fits of the ETV curves yielded $dP/dt=-1.196\pm0.003\times10^{-5}$\,d/yr for KIC 9840412 and $-2.737\pm0.089\times10^{-7}$\,d/yr for KIC 10292413. Follow-up observations of these two systems (especially minima timing) are recommended in order to confirm or rule out LTTE as the cause for ETV and to establish whether they can be considered as stellar merger candidates.

The Optical Gravitational Lensing Experiment (OGLE) has also been a reliable source for long-term studies of eclipsing binaries. The period changes of 569 OGLE contact binaries were analysed by Kubiak et al. (2006). Notably, the two systems with the most rapid period decrease have very long periods: 1.36\,d and 1.70\,d respectively. The error of calculation of period changes is assessed at $2.3\times10^{-7}$\,d/yr by the authors. All 14 systems with $P>0.8$\,d and $dP/dt<-2.3\times10^{-7}$\,d/yr are listed in Tab.\,\ref{ogle}. They require a large-sized telescope for follow-up observations considering that they are faint objects in a crowded field near the galactic center.

\begin{table}
\begin{center}
\caption[A selection of long-period OGLE contact binaries with decreasing periods (Kubiak et al. 2006).]{A selection of long-period OGLE contact binaries with decreasing periods (Kubiak et al. 2006).}\label{ogle}
\small
\begin{tabular}{c@{ }c@{ }c@{ }c@{ }c@{ }c@{ }}
\hline\hline
$\alpha$ (J2000) \,& $\delta$ (J2000)  \,& $I$ [mag] \,& $V-I$ \,& P [d] \,& dP/dt [d/yr] \\
\hline
18:01:57.16 \,&  -30:17:34.6 \,&  16.54 \,& 1.37  \,&	1.3642907  \,&   -4.900E-06	 \\
18:01:15.04 \,&  -29:50:50.7 \,&  17.74 \,& 1.63  \,&	1.7027067  \,&   -3.700E-06	 \\
18:03:59.26 \,&  -29:59:41.2 \,&  16.07 \,& 1.11  \,&	1.2046764  \,&   -2.300E-06	 \\
18:02:44.75 \,&  -30:19:51.9 \,&  15.85 \,& 1.26  \,&	1.0911776  \,&   -2.100E-06	 \\
18:02:44.24 \,&  -29:48:45.4 \,&  15.55 \,& 1.16  \,&	0.8556117  \,&   -1.700E-06	 \\
18:03:53.87 \,&  -30:19:19.8 \,&  16.75 \,& 1.26  \,&	0.8084745  \,&   -1.600E-06	 \\
18:01:52.74 \,&  -30:18:17.9 \,&  16.85 \,& 1.37  \,&	1.0291539  \,&   -1.400E-06	 \\
18:02:24.11 \,&  -30:07:38.7 \,&  17.10 \,& 1.42  \,&	0.8101012  \,&   -1.300E-06	 \\
18:01:59.16 \,&  -29:45:34.4 \,&  15.98 \,& 1.28  \,&	0.8924708  \,&   -8.300E-07	 \\
18:02:37.56 \,&  -30:20:17.7 \,&  17.38 \,& 1.33  \,&	0.8552266  \,&   -8.100E-07	 \\
18:03:21.07 \,&  -29:59:32.3 \,&  16.41 \,& 1.12  \,&	0.9480481  \,&   -7.300E-07	 \\
18:03:56.25 \,&  -29:49:46.0 \,&  16.61 \,& 1.33  \,&	0.8890325  \,&   -4.600E-07	 \\
18:02:13.56 \,&  -29:48:37.8 \,&  16.70 \,& 1.30  \,&	0.9320552  \,&   -4.000E-07	 \\
18:01:50.77 \,&  -29:52:46.0 \,&  18.05 \,& 1.20  \,&	0.8133902  \,&   -3.100E-07	 \\
\hline        
\end{tabular} 
\end{center}  
\end{table}

\section*{5. Summary and discussion}

The only red nova with a well-studied progenitor is V1309 Sco, which erupted in 2008. Its progenitor was a contact binary with an exceptionally long period of $\sim1.44$\,d, which could be explained by the evolutionary state of the system. The present work aims to explore the current prospects of searching for similar systems and stellar merger candidates as a whole. 

The AAVSO VSX database was used to select periodic variables classified as contact binaries. Of the systems in the $1.1-1.8$\,d period range 207 were identified in the SDSS catalog. By utilizing the color-temperature relations derived by Boyajian et al. (2013) approximate surface temperatures for 128 systems were obtained. In 15 cases the calculated temperatures are in the $4200-4800$\,K range and the maximum possible reddening is lower than $0.07$\,mag. If correctly classified as W UMa-type variables, these 15 systems may have similar parameters as the V1309 Sco progenitor. Therefore, a more detailed exploration of these systems could bring an insight into the physics of a red nova progenitor.

An prediction of a stellar merger in the near future would require evidence that the orbital period is decreasing. A steady decrease of the orbital period results in a parabolic ETV curve with a negative second derivative. The 31 Kepler systems with parabolic ETV curves selected by Conroy et al. (2014) were visually classified. Two of them were selected as potential stellar merger candidates and require further observations. A search among the fainter OGLE contact binaries yielded 14 systems with $P>0.8$\,d and a steady period decrease.

The number of known eclipsing binaries is still insufficient to select exact twins of the V1309 Sco progenitor. However, it is already sufficient to select systems with similar parameters and, possibly, in a similar evolutionary state. Some of those also have apparent magnitudes comparable to the V1309 Sco progenitor (Tab.\,\ref{results}). Upcoming missions such as GAIA and LSST are expected to discover millions of eclipsing binaries (e.g. Pr{\v s}a et al. 2011). Therefore the number of known contact binaries with evolved components may increase by a factor of $\sim10^{2}$ in the following decade.


\newpage

\begin{table}[ht!]
\begin{center}
\caption[.]{Photospheric temperatures of 128 contact binaries with periods in the $1.1-1.8$\,d range and galactic latitudes $|b|>20$\degr.}\label{all}
\scriptsize
\begin{tabular}{c@{ }c@{ }c@{ }c@{ }c@{ }c@{ }c@{ }c@{ }c@{ }c@{ }c@{ }c@{ }c@{ }c@{ }}
\hline\hline
\TT $\alpha$(J2000.0) \,\,&  $\delta$(J2000.0)  \,\,&  P [d] \,\,&  Mag. \,\,&  Amp. \,\,&  $T_{eff}$[K] &  $\sigma_{T}$[K] \,&  $\alpha$(J2000.0) \,\,&  $\delta$(J2000.0)  \,\,&  P [d] \,\,&  Mag. \,\,&  Amp. \,\,&  $T_{eff}$[K] &  $\sigma_{T}$[K]  \\
\hline
   00:24:35.21 \,&   07:34:25.5 \,&   1.7100 \,&    13.92 \,&    0.08 \,&     3306 \,&    132   \,&    13:02:13.45 \,&   34:05:07.2 \,&   1.6378 \,&	16.57 \,&    0.34 \,&	  5801 \,&     61   \\
   00:34:56.35 \,&   24:15:04.9 \,&   1.5581 \,&    17.21 \,&    0.28 \,&     6310 \,&     93   \,&    13:03:46.26 \,&   00:46:43.2 \,&   1.3072 \,&	16.68 \,&    0.66 \,&	  5023 \,&     81   \\
   00:44:12.88 \,&   20:11:07.8 \,&   1.7533 \,&    13.75 \,&    0.06 \,&     4750 \,&     77   \,&    13:06:00.14 \,&   18:00:46.7 \,&   1.3074 \,&	16.68 \,&    0.14 \,&	  4315 \,&     70   \\
   00:49:18.55 \,&   39:52:29.7 \,&   1.1784 \,&    15.93 \,&    0.36 \,&     6354 \,&     39   \,&    13:24:23.64 \,&   07:58:59.8 \,&   1.6895 \,&	16.91 \,&    0.19 \,&	  6561 \,&    118   \\
   00:58:44.79 \,&   18:43:29.7 \,&   1.2727 \,&    17.38 \,&    0.23 \,&     4103 \,&     42   \,&    14:07:20.29 \,&  -12:45:15.9 \,&   1.4548 \,&	15.10 \,&    0.35 \,&	  6563 \,&     81   \\
   01:02:35.85 \,&   40:27:47.4 \,&   1.5630 \,&    17.70 \,&    0.46 \,&     5233 \,&     53   \,&    14:18:14.46 \,&   41:36:41.8 \,&   1.3134 \,&	16.09 \,&    0.22 \,&	  6236 \,&     51   \\
   01:04:25.72 \,&  -05:03:30.3 \,&   1.2006 \,&    15.04 \,&    0.18 \,&     3891 \,&     18   \,&    14:21:14.77 \,&   35:28:38.9 \,&   1.5851 \,&	16.14 \,&    0.09 \,&	  4592 \,&     26   \\
   01:56:09.97 \,&   28:40:18.7 \,&   1.2016 \,&    14.22 \,&    0.12 \,&     7415 \,&     81   \,&    14:25:27.50 \,&   30:41:49.9 \,&   1.1985 \,&	17.91 \,&    0.41 \,&	  3539 \,&     71   \\
   02:09:03.05 \,&  -04:26:14.7 \,&   1.5781 \,&    18.17 \,&    0.31 \,&     4496 \,&    116   \,&    14:27:59.50 \,&   06:54:24.0 \,&   1.7529 \,&	15.50 \,&    0.09 \,&	  4937 \,&     48   \\
   02:23:29.59 \,&   11:57:23.6 \,&   1.2330 \,&    14.66 \,&    0.17 \,&     4864 \,&    100   \,&    15:02:20.15 \,&   26:21:12.1 \,&   1.7954 \,&	15.94 \,&    0.18 \,&	  4729 \,&     83   \\
   02:31:59.75 \,&  -03:58:32.3 \,&   1.4094 \,&    14.62 \,&    0.11 \,&     3373 \,&    100   \,&    15:13:48.47 \,&   12:32:11.0 \,&   1.5098 \,&	16.69 \,&    0.10 \,&	  5093 \,&     61   \\
   02:48:49.65 \,&   19:10:16.0 \,&   1.6974 \,&    14.34 \,&    0.18 \,&     5721 \,&    130   \,&    15:20:53.17 \,&  -13:23:15.7 \,&   1.7053 \,&	14.78 \,&    0.19 \,&	  5987 \,&    147   \\
   02:56:11.59 \,&   32:42:52.4 \,&   1.2569 \,&    13.40 \,&    0.28 \,&     5340 \,&     55   \,&    15:35:10.57 \,&   01:21:19.1 \,&   1.5204 \,&	16.13 \,&    0.69 \,&	  5369 \,&    109   \\
   03:04:26.92 \,&   18:36:33.2 \,&   1.3890 \,&    17.39 \,&    0.44 \,&     6385 \,&    134   \,&    15:35:46.16 \,&   17:36:01.8 \,&   1.3456 \,&	13.97 \,&    0.08 \,&	  5453 \,&     98   \\
   03:18:05.55 \,&   17:55:27.0 \,&   1.1015 \,&    14.78 \,&    0.09 \,&     5591 \,&     54   \,&    15:36:08.42 \,&   21:09:38.2 \,&   1.5163 \,&	18.22 \,&    0.46 \,&	  6402 \,&     34   \\
   03:23:29.17 \,&  -00:34:49.3 \,&   1.7183 \,&    15.66 \,&    0.15 \,&     5152 \,&     76   \,&    15:39:45.08 \,&   07:20:46.2 \,&   1.3350 \,&	16.21 \,&    0.26 \,&	  6802 \,&     73   \\
   03:54:59.63 \,&   01:16:16.3 \,&   1.5202 \,&    15.89 \,&    0.32 \,&     3631 \,&    111   \,&    15:50:21.58 \,&   12:22:16.2 \,&   1.1158 \,&	17.48 \,&    0.50 \,&	  6583 \,&     94   \\
   04:28:43.41 \,&  -06:53:22.3 \,&   1.3491 \,&    13.99 \,&    0.15 \,&     3618 \,&    140   \,&    15:51:53.35 \,&   04:15:29.8 \,&   1.4075 \,&	14.36 \,&    0.14 \,&	  6060 \,&     85   \\
   04:45:50.71 \,&   11:34:57.2 \,&   1.3164 \,&    15.29 \,&    0.35 \,&     4723 \,&    110   \,&    16:06:30.09 \,&   33:13:44.1 \,&   1.5736 \,&	14.81 \,&    0.11 \,&	  6257 \,&     64   \\
   04:52:13.32 \,&   00:56:18.4 \,&   1.6971 \,&    14.68 \,&    0.09 \,&     6638 \,&     33   \,&    16:15:39.64 \,&   12:47:23.2 \,&   1.1409 \,&	16.15 \,&    0.24 \,&	  5104 \,&    101   \\
   04:56:06.10 \,&  -01:07:53.1 \,&   1.1575 \,&    14.56 \,&    0.18 \,&     3672 \,&     62   \,&    16:20:45.25 \,&   34:57:41.6 \,&   1.2987 \,&	16.98 \,&    0.13 \,&	  6468 \,&     63   \\
   05:06:27.90 \,&  -00:21:30.0 \,&   1.2615 \,&    15.59 \,&    0.10 \,&     3382 \,&     22   \,&    16:26:09.66 \,&   05:13:20.5 \,&   1.4847 \,&	15.89 \,&    0.13 \,&	  4102 \,&     44   \\
   05:27:06.19 \,&  -05:58:27.6 \,&   1.4527 \,&    15.74 \,&    0.11 \,&     3283 \,&     58   \,&    16:29:45.49 \,&  -02:11:41.6 \,&   1.5380 \,&	17.38 \,&    0.33 \,&	  5377 \,&     97   \\
   07:26:18.08 \,&   31:55:03.0 \,&   1.1579 \,&    15.71 \,&    0.22 \,&     4744 \,&     67   \,&    16:30:29.76 \,&   20:23:16.4 \,&   1.6452 \,&	17.40 \,&    0.42 \,&	  6001 \,&    139   \\
   07:34:32.94 \,&   36:48:28.5 \,&   1.1646 \,&    17.38 \,&    0.31 \,&     4519 \,&     96   \,&    16:30:42.51 \,&   53:12:58.7 \,&   1.1418 \,&	17.59 \,&    0.34 \,&	  5588 \,&     69   \\
   07:38:45.77 \,&   34:57:20.3 \,&   1.5792 \,&    13.30 \,&    0.17 \,&     5481 \,&     93   \,&    16:41:42.79 \,&   20:20:37.2 \,&   1.1087 \,&	15.14 \,&    0.10 \,&	  6123 \,&     60   \\
   07:39:47.77 \,&   32:40:24.6 \,&   1.5975 \,&    13.31 \,&    0.14 \,&     5390 \,&     32   \,&    16:52:47.42 \,&   16:18:35.1 \,&   1.6941 \,&	13.84 \,&    0.19 \,&	  4130 \,&     75   \\
   08:05:06.44 \,&   14:51:38.3 \,&   1.3981 \,&    17.58 \,&    0.26 \,&     6126 \,&     41   \,&    16:56:09.74 \,&   42:42:19.7 \,&   1.2699 \,&	15.52 \,&    0.18 \,&	  4383 \,&    127   \\
   08:05:43.15 \,&   40:46:38.5 \,&   1.5851 \,&    15.40 \,&    0.08 \,&     4908 \,&     37   \,&    17:00:41.65 \,&   12:05:31.6 \,&   1.5080 \,&	14.65 \,&    0.11 \,&	  6036 \,&     83   \\
   08:11:08.14 \,&   30:38:03.3 \,&   1.6462 \,&    14.48 \,&    0.06 \,&     4105 \,&     92   \,&    17:10:08.19 \,&   32:56:11.3 \,&   1.2843 \,&	17.80 \,&    0.36 \,&	  5398 \,&     56   \\
   08:17:46.29 \,&   13:05:05.7 \,&   1.5947 \,&    14.90 \,&    0.08 \,&     4519 \,&     64   \,&    17:12:06.21 \,&   40:55:37.8 \,&   1.5102 \,&	15.77 \,&    0.16 \,&	  5504 \,&     60   \\
   08:19:42.20 \,&   41:29:47.9 \,&   1.6497 \,&    16.13 \,&    0.14 \,&     3743 \,&     74   \,&    17:14:28.01 \,&   10:13:10.7 \,&   1.7623 \,&	15.92 \,&    0.14 \,&	  4479 \,&     61   \\
   08:22:08.52 \,&   09:07:27.1 \,&   1.2965 \,&    17.93 \,&    0.31 \,&     6077 \,&     82   \,&    17:18:55.47 \,&   62:03:28.6 \,&   1.1993 \,&	14.15 \,&    0.40 \,&	  6800 \,&     28   \\
   08:29:50.38 \,&   23:06:51.8 \,&   1.6032 \,&    16.07 \,&    0.10 \,&     6935 \,&     32   \,&    17:20:10.09 \,&   08:26:36.7 \,&   1.6102 \,&	15.41 \,&    0.42 \,&	  5945 \,&    121   \\
   08:31:48.57 \,&   29:44:44.0 \,&   1.4136 \,&    14.93 \,&    0.12 \,&     4445 \,&     89   \,&    17:25:53.14 \,&   61:17:15.6 \,&   1.1309 \,&	14.41 \,&    0.27 \,&	  7551 \,&	9   \\
   08:43:34.88 \,&   03:09:57.0 \,&   1.3806 \,&    16.47 \,&    0.10 \,&     3970 \,&     13   \,&    17:27:21.85 \,&   27:50:55.3 \,&   1.5273 \,&	18.01 \,&    0.44 \,&	  5283 \,&     45   \\
   08:45:36.01 \,&   20:13:00.6 \,&   1.3800 \,&    17.00 \,&    0.11 \,&     6297 \,&     88   \,&    17:35:10.22 \,&   20:31:02.2 \,&   1.2142 \,&	16.37 \,&    0.18 \,&	  5645 \,&     75   \\
   08:49:23.10 \,&   10:51:29.1 \,&   1.1003 \,&    16.43 \,&    0.10 \,&     4148 \,&     24   \,&    17:45:56.93 \,&   42:24:00.5 \,&   1.2803 \,&	16.52 \,&    0.10 \,&	  6162 \,&     81   \\
   08:58:13.10 \,&   13:38:20.5 \,&   1.4357 \,&    14.95 \,&    0.12 \,&     4886 \,&    101   \,&    20:53:56.62 \,&  -01:09:43.6 \,&   1.6379 \,&	14.14 \,&    0.17 \,&	  5021 \,&     65   \\
   09:01:32.24 \,&   01:01:01.8 \,&   1.5418 \,&    16.27 \,&    0.10 \,&     6112 \,&     83   \,&    21:00:34.55 \,&  -11:38:46.2 \,&   1.1281 \,&	17.91 \,&    0.45 \,&	  6798 \,&    148   \\
   09:03:58.49 \,&   16:17:12.7 \,&   1.2009 \,&    16.72 \,&    0.17 \,&     5154 \,&     80   \,&    21:15:59.21 \,&  -10:50:10.4 \,&   1.3021 \,&	14.64 \,&    0.13 \,&	  6256 \,&     77   \\
   09:05:04.20 \,&   23:44:07.1 \,&   1.3327 \,&    16.13 \,&    0.91 \,&     6396 \,&     85   \,&    21:48:28.60 \,&   09:03:36.5 \,&   1.4246 \,&	16.87 \,&    0.17 \,&	  5994 \,&    131   \\
   09:06:32.35 \,&   11:29:30.8 \,&   1.1806 \,&    15.83 \,&    0.13 \,&     4995 \,&     86   \,&    21:48:47.47 \,&  -05:48:26.1 \,&   1.5972 \,&	17.57 \,&    0.22 \,&	  6200 \,&     90   \\
   09:07:14.17 \,&   13:27:20.7 \,&   1.1835 \,&    16.97 \,&    0.36 \,&     7224 \,&     51   \,&    21:50:41.97 \,&   12:37:08.2 \,&   1.2976 \,&	15.94 \,&    0.09 \,&	  4000 \,&     27   \\
   09:15:04.25 \,&   29:52:40.1 \,&   1.3513 \,&    14.39 \,&	 0.28 \,&     5288 \,&     18	\,&    21:58:23.25 \,&   07:18:31.5 \,&   1.2970 \,&	16.52 \,&    0.11 \,&	  6116 \,&     75   \\
   09:46:14.73 \,&   16:47:06.0 \,&   1.6007 \,&    15.31 \,&	 0.25 \,&     4750 \,&    100	\,&    22:00:17.17 \,&   23:21:32.0 \,&   1.6262 \,&	15.04 \,&    0.43 \,&	  4948 \,&     48   \\
   09:46:56.35 \,&   40:14:59.8 \,&   1.4679 \,&    14.67 \,&	 0.23 \,&     5378 \,&     84	\,&    22:07:38.21 \,&   30:03:53.1 \,&   1.1424 \,&	14.65 \,&    0.37 \,&	  5292 \,&     73   \\
   09:50:00.13 \,&   23:49:52.2 \,&   1.1698 \,&    15.44 \,&	 0.09 \,&     3970 \,&     21	\,&    22:16:20.77 \,&   07:33:57.2 \,&   1.1351 \,&	15.60 \,&    0.07 \,&	  4026 \,&     15   \\
   09:54:04.15 \,&   05:56:33.0 \,&   1.3438 \,&    16.05 \,&	 0.12 \,&     3796 \,&     35	\,&    22:29:09.15 \,&   17:37:23.4 \,&   1.3664 \,&	14.60 \,&    0.08 \,&	  4568 \,&     46   \\
   10:00:01.70 \,&   24:23:05.1 \,&   1.2955 \,&    15.85 \,&	 0.15 \,&     4213 \,&     48	\,&    22:32:25.40 \,&   06:00:48.3 \,&   1.4293 \,&	16.12 \,&    0.11 \,&	  5879 \,&    104   \\
   10:02:30.35 \,&   50:18:08.8 \,&   1.2404 \,&    15.51 \,&	 0.23 \,&     4760 \,&     74	\,&    22:33:22.32 \,&   30:33:25.0 \,&   1.2756 \,&	15.90 \,&    0.24 \,&	  5848 \,&    105   \\
   10:38:20.36 \,&   12:46:14.9 \,&   1.3373 \,&    15.72 \,&	 0.16 \,&     3267 \,&     84	\,&    22:42:01.99 \,&  -05:25:00.1 \,&   1.7298 \,&	16.58 \,&    0.48 \,&	  5144 \,&     75   \\
   10:38:44.77 \,&   13:45:51.1 \,&   1.3263 \,&    14.63 \,&	 0.10 \,&     3456 \,&     94	\,&    22:42:18.68 \,&   14:24:09.1 \,&   1.4598 \,&	16.37 \,&    0.11 \,&	  3452 \,&     64   \\
   10:49:39.27 \,&   33:45:22.1 \,&   1.3964 \,&    14.55 \,&	 0.41 \,&     5217 \,&     52	\,&    22:43:42.38 \,&   26:02:16.9 \,&   1.1415 \,&	15.13 \,&    0.19 \,&	  6710 \,&     16   \\
   11:00:13.96 \,&   36:04:29.0 \,&   1.6927 \,&    16.30 \,&	 0.32 \,&     6793 \,&     14	\,&    22:45:43.47 \,&   07:47:14.2 \,&   1.3456 \,&	15.68 \,&    0.09 \,&	  5147 \,&     93   \\
   11:07:43.59 \,&   21:13:51.7 \,&   1.2885 \,&    17.67 \,&	 0.24 \,&     6167 \,&     40	\,&    22:56:43.91 \,&  -02:56:22.3 \,&   1.1710 \,&	15.51 \,&    0.10 \,&	  3533 \,&     54   \\
   11:08:58.93 \,&   31:45:52.3 \,&   1.2302 \,&    16.07 \,&	 0.37 \,&     5625 \,&    106	\,&    23:08:23.17 \,&   15:17:11.5 \,&   1.5196 \,&	17.39 \,&    0.37 \,&	  4804 \,&     80   \\
   11:11:01.12 \,&   03:31:00.9 \,&   1.5349 \,&    14.49 \,&	 0.41 \,&     6009 \,&     75	\,&    23:12:46.22 \,&   17:13:14.5 \,&   1.1487 \,&	14.92 \,&    0.37 \,&	  6403 \,&     56   \\
   11:16:12.89 \,&   00:17:52.6 \,&   1.4245 \,&    16.13 \,&	 0.51 \,&     5017 \,&     71	\,&    23:15:47.32 \,&   12:29:37.5 \,&   1.7325 \,&	15.28 \,&    0.19 \,&	  4679 \,&     75   \\
   11:18:17.29 \,&  -00:14:45.2 \,&   1.4907 \,&    15.95 \,&	 0.75 \,&     5875 \,&     54	\,&    23:22:22.15 \,&   22:06:22.3 \,&   1.6478 \,&	17.02 \,&    0.16 \,&	  5218 \,&     42   \\
   12:10:22.21 \,&   35:46:55.4 \,&   1.5242 \,&    17.86 \,&	 0.19 \,&     6231 \,&    104	\,&    23:33:31.70 \,&   14:37:43.1 \,&   1.1161 \,&	17.37 \,&    0.70 \,&	  4884 \,&     75   \\
   12:30:23.09 \,&  -02:23:36.7 \,&   1.3086 \,&    16.38 \,&	 0.11 \,&     6382 \,&     91	\,&    23:36:48.71 \,&   35:10:36.9 \,&   1.1653 \,&	14.53 \,&    0.15 \,&	  3987 \,&     26   \\
   12:32:25.24 \,&   04:02:34.8 \,&   1.3214 \,&    16.43 \,&	 0.15 \,&     6277 \,&     54	\,&    23:54:34.83 \,&   35:28:09.3 \,&   1.7990 \,&	14.41 \,&    0.16 \,&	  5706 \,&    121   \\
   12:48:15.08 \,&   13:14:21.4 \,&   1.2449 \,&    17.75 \,&	 0.27 \,&     6171 \,&     78	\,&    23:55:46.62 \,&   29:36:20.6 \,&   1.1607 \,&	15.91 \,&    0.57 \,&	  5120 \,&     89   \\
\hline        
\end{tabular} 
\end{center}  
\end{table}


\begin{thebibliography}{}

\bibitem[1]{2012ApJS..203...21A}
Ahn C. P., Alexandroff R., Allende Prieto C. et al., 2012, {\em \apjs}, 203, 21

\bibitem[2]{2003Natur.422..405B}
Bond H. E., Henden A., Levay Z. G. et al., 2003, {\em \nat}, 422, 405-408

\bibitem[3]{2013ApJ...771...40B}
Boyajian T. S., von Braun K., van Belle G. et al., 2013, {\em \apj}, 771, 40

\bibitem[4]{2014AJ....147...45C}
Conroy K. E., Pr{\v s}a A., Stassun K. G. et al., 2014, {\em \aj}, 147, 45

\bibitem[5]{2014ApJS..213....9D}
Drake A. J., Graham M. J., Djorgovski S. G. et al., 2014, {\em \apjs}, 213, 9

\bibitem[6]{2015Natur.520..322K}
Kami\'nski T., Menten K. M., Tylenda R. et al., 2015, {\em \nat}, 520, 322-324

\bibitem[7]{2014MNRAS.443.1319K}
Kochanek C. S., Adams S. M. and Belczynski K., 2014, {\em \mnras}, 443, 1319-1328

\bibitem[8]{2006AcA....56..253K}
Kubiak M., Udalski A. and Szymanski M. K., 2006, {\em AcA}, 56, 253

\bibitem[9]{2007Natur.447..458K}
Kulkarni S. R., Ofek E. O., Rau A. et al., 2007, {\em \nat}, 447, 458-460

\bibitem[10]{2015A&A...578L..10K}
Kurtenkov A. A., Pessev P., Tomov T. et al., 2015, {\em \aap}, 578, L10

\bibitem[11]{2010A&A...516A.108M}
Mason E., Diaz M., Williams R. E., Preston G. and Bensby T., 2010, {\em \aap}, 516, A108

\bibitem[12]{2015AAS...22541505M}
Molnar L. A., Van Noord D. M., Steenwyk S. D., Spedden C. J. and Kinemuchi K., 2010, {\em AAS Meeting Abstracts}, 225, 415.05

\bibitem[13]{2011AJ....142...52P}
Pr{\v s}a A., Pepper J. and Stassun K. G., 2011, {\em \aj}, 142, 52

\bibitem[14]{1989ApJ...341L..51R}
Rich R. M., Mould J., Picard A., Frogel J. A. and Davies R., 1989, {\em \apjl}, 341, L51-L54

\bibitem[15]{2011AJ....142..160S}
Slawson R. W., Pr{\v s}a A., Welsh W. F. et al., 2011, {\em \aj}, 142, 160

\bibitem[16]{2011A&A...528A.114T}
Tylenda R., Hajduk M., Kami{\'n}ski et al., 2011, {\em \aap}, 528, A114

\bibitem[17]{2006A&A...451..223T}
Tylenda R. and Soker N., 2006, {\em \aap}, 451, 223-236

\end{thebibliography}
\end{document}